%% file: main.tex
\documentclass{article}
\usepackage{style}

\title{1PN effective binary Lagrangian for the gravity-Kalb-Ramond sector in the conservative regime}
\author{Vegard Undheim}
\author{Alex B. Nielsen}
\author{Eirik Eik Svanes}
\affil{Department of Mathematics and Physics, University of Stavanger, Norway}

 \date{\today}

\begin{document}

\maketitle

\begin{abstract}
     Within the framework of string theory, a number of new fields are possible correcting the Einstein-Hilbert action, including a Kalb-Ramond two-form field. In this work we derive explicitly first order relativistic corrections to conservative dynamics with a Kalb-Ramond field, using the effective field theory approach. The resulting additional terms in the Lagrangian governing conservative binary dynamics are presented explicitly.
\end{abstract}

\section{Introduction}

String theory postulates a theory of quantum gravity, which generically includes extra dimensions. This leads to modifications of Einstein's theory of general relativity. However, it is not known at which scale these modifications will manifest themselves, and this depends heavily on which ``string compactification" is employed.  At lower energies gravitational effects in the extra dimensions will give rise to new fields called moduli fields, generically present in most models derived from string theory. These fields, and in particular their axionic components and the Peccei–Quinn axion \cite{Peccei:1977hh, Wilczek:1977pj, Weinberg:1977ma, Kim:1979if, Vafa:1984xg, GrillidiCortona:2015jxo}, have been suggested as natural candidates for light dark matter, where a plethora of physical models and searches have been suggested for their observations \cite{Svrcek:2006yi, Arvanitaki:2009fg, Marsh:2015xka}. 
The effects of light particles and QCD axions in gravitational mergers have also been studied in the literature before, see for example \cite{AxionLIGO}.

In addition to moduli fields, string theory and more generally supergravity theories, are equipped with a Kalb-Ramond two-form field \cite{Kalb:1974yc}. This field was originally introduced as a classical generalisation of the electromagnetic interaction of point particles to one-dimensional strings in the Feynman-Wheeler picture \cite{Kalb:1974yc}.  The Kalb-Ramond field is now understood as a generic feature of supergravity theories, forming part of the gravitational supermultiplet.\footnote{Here, we are explicitly focussing only on contributions from the Kalb-Ramond field, ignoring the dilaton and other supergravity effects. Generic four-dimensional solutions with such fields turned on were found in \cite{Burgess:1994kq, Kang:2019owv}.} Classically upon dualising, the Kalb-Ramond field may be thought of as a ``fundamental string axion'' \cite{Svrcek:2006yi}, although the physical and quantum behaviours of this fundamental axion are slightly different to the QCD axion. For example, the fundamental Kalb-Ramond axion is required to be massless at the level of the classical supergravity action, as its coupling is dictated by classical gauge symmetry. 

In string theory the Kalb-Ramond two-form is responsible for mediating the force between fundamental strings, whose natural energy scale is the string scale. Being a massless mediator, the classical force is expected to be long range. However, as the couplings are string scale suppressed, any direct measurement of such interactions seem a long way off. The recent discoveries by NANOGrav are however interesting in this regard, as cosmic superstrings provide a potential explanation of the data \cite{NANOGrav:2023hvm}. As fundamental strings, these would then also provide large sources of potentially detectable Kalb-Ramond charge \cite{Copeland:2003bj}.

Another possibility is that a large number of strings can somehow amalgamate to produce a compound effect which can be measured. One such scenario where this can potentially happen is in the case of fuzzballs, see \cite{Mathur:2005zp} and references therein. This string-inspired alternative to black holes has a large number of strings condense at the horizon of the black hole, where space time effectively ends. The accumulated Kalb-Ramond charge may then have a chance of being observed, and it is this coupling we wish to model in the present paper. We'll do this by classically modelling the Kalb-Ramond field as an axionic scalar, where astrophysical objects can be charged under this field. At the quantum level, the Kalb-Ramond field should be treated as a two-form, whose interactions are rather more subtle and involved. We will leave the quantum analysis for future work. 

In this work, we derive classical corrections to the gravitational potential for a binary system. Recent years have seen substantial improvements in constraining binary motion for relativistic systems. One of these is the pulsar timing with radio telescopes of the double pulsar \cite{Kramer:2021jcw}. Another is the constraints from gravitational wave signals from binary inspiral \cite{LIGOScientific:2019fpa}. These are expected to improve by orders of magnitude over the coming decades \cite{KAGRA:2013rdx,Maggiore:2019uih,Barausse:2020rsu}. 

To derive the corrections due to the Kalb-Ramond field, we employ here the effective field theory formalism \cite{Goldberger:EFT,Goldberger:LesHouches}. This formalism allows an explicit separation of scales and has been used in a number of other studies, see e.g. \cite{AxionLIGO} and others. Our work here serves to illustrate the utility of this formalism for studying additional fields from particle physics and including them in gravitational interactions. For completeness we also explicitly include pure graviton interactions at the same order. As noted, we restrict ourselves to tree-level computations and do not include quantum effects in our analysis. Our main result is equation \eqref{eq:Lagragian:KR}. This differs qualitatively from previous analysis involving axionic scalars \cite{AxionLIGO}. In particular, the Kalb-Ramond and classical General Relativity corrections have very similar $\tfrac{1}{r}$-behaviours in the $\tfrac{v}{c}$ expansion. We speculate how this relates to understanding the Kalb-Ramond field in the context of generalised geometry,\footnote{For an introduction to the mathematical framework of generalised geometry, see \cite{hitchin2010lectures}, and for applications in supergravity, see e.g. \cite{Coimbra:2011nw, Coimbra:2012af} with references therein.} where the metric and Kalb-Ramond field are put on the same footing in the generalised metric.

\input{KRcalculations}

\section{Conclusion and Outlook}
    We have derived the leading order relativistic Lagrangian corrections for a Kalb-Ramond field coupled to gravity, Eqn.(\ref{eq:Lagragian:KR}). Our computation would apply in the relativistic regime, but not the quantum regime. Astrophysical compact objects may or may not have measurable Kalb-Ramond charge. In the stringy fuzzball proposal, black holes are composed of macroscopic strings \cite{Mathur:2005zp} and these may well have non-zero Kalb-Ramond charge. Fuzzballs should potentially also have non-trivial dilation profile, as happens in other black-hole solutions in dilaton gravity and supergravity \cite{Burgess:1994kq, Kang:2019owv, Abedi:2022omf}. But unlike the Kalb-Ramond (and photon) field, the dilaton is not protected by a symmetry and so can dynamically acquire a mass, leading to a short range force even at the classical level. If macroscopic strings exist at cosmological scales, then they may also interact via a Kalb-Ramond mediated force \cite{Copeland:2003bj}.


    In our derived Lagrangian, note the similar $\tfrac{1}{r}$-behaviours in the relativistic expansions for both the Kalb-Ramond and classical General Relativity corrections. This is perhaps not too surprising in view of generalised geometry, where the metric and Kalb-Ramond fields are put on the same footing in the generalised metric \cite{hitchin2010lectures, Coimbra:2011nw, Coimbra:2012af}. This point of view also leads us to speculate that this similar behaviour will persist to higher orders. In order to investigate this further, it would be very useful to extend the perturbative Feynman diagram formalism of \cite{Goldberger:EFT,Goldberger:LesHouches} to the generalised framework. This framework will also be relevant when we come to understand the theory in higher dimensions or with quantum corrections, where the Kalb-Ramond field cannot be treated as a scalar anymore.     
    
    The approach presented here also demonstrates the power of the effective field theory approach to classifying and calculating possible deviations from Einstein's theory of gravity. This is relevant in an era of precision pulsar and gravitational wave observations that are testing Einstein's theory in the strongly relativistic regime.


\section*{Acknowledgements}
We thank Jahed Abedi and Edward Hardy for useful conversations.

\printbibliography
\end{document}

%% file: KRcalculations.tex
\section{Effective action for the Kalb-Ramond field}

Our aim is to calculate the first-order relativistic correction to the interaction of two bodies charged under the Kalb-Ramond field. We therefore study a model in which the Kalb-Ramond field is minimally coupled to gravity alone.\bigskip

\noindent The Kalb-Ramond field $B_{\mu\nu}$ is a two-form on spacetime with a field strength $H\ind{_{\mu\nu\rho}}$ given by
\beq
H\ind{_{\mu\nu\rho}} \equiv \dd B\indices{_{\mu\nu\rho}} = B\indices{_{\nu\rho,\mu}} + B\indices{_{\rho\mu,\nu}} + B\indices{_{\mu\nu,\rho}} \label{eq:EoM:H:Exact}
\eeq
The action and equations of motion (EoM) for the metric including the Kalb-Ramond field are\footnote{We here define the action as $S = \int L \dd{t} = \int \mathcal{L} \dd[3]{x} \dd{t}$, thus we may write $\int \mathcal{L} \dd[4]{x} = cS $ for notational convenience.}
\begin{align}
    & cS = \int \dd[4]{x} \sqrt{-g} \left(\frac{1}{2\kappa} R + \frac{1}{12} H\indices{_{\mu\nu\rho}} H\indices{^{\mu\nu\rho}} + \mathcal{L}_\text{matter} \right), \label{eq:Action:full} \\
    & \nabla\ind{_\sigma} H\indices{^\sigma_{\mu\nu}} = 0, \label{eq:EoM:H} 
\end{align}

\subsection{A solution for the Kalb-Ramond field}
The solution for the $H$-field, equation \eqref{eq:EoM:H}, has been known for a long time, and can be found in \cite{Kalb:1974yc}. Since $H\ind{_{\mu\nu\rho}}$ is \emph{exact} by equation \eqref{eq:EoM:H:Exact}, we make the ansatz that $H\ind{_{\mu\nu\rho}}$ can be described by the derivative of a scalar function with an antisymmetric tensor structure, such that
\begin{align}
    H\indices{_{\alpha\beta\gamma}} = \varepsilon\indices{_{\alpha\beta\gamma\delta}} K\indices{^{;\delta}}\:, \label{eq:sol:H}
\end{align}
where $\varepsilon\indices{_{\alpha\beta\gamma\delta}} = \sqrt{-\abs{g}} \epsilon\indices{_{\alpha\beta\gamma\delta}}$ is the totally anti-symmetric volume form of four space-time dimensions, i.e. the Levi-Civita tensor.
This is a solution of equation \eqref{eq:EoM:H} based on the antisymmetry of the indices $\alpha \beta \gamma \delta$, and the symmetry of the derivative indices $\alpha \delta$: $\nabla_\alpha H\indices{^{\alpha}_{\beta\gamma}}=\nabla_\alpha \nabla_\delta \varepsilon\indices{^\alpha_{\beta\gamma}^\delta} K$.

The anzats \eqref{eq:sol:H} can be further constrained by making sure $H\indices{_{\alpha\beta\gamma}}$ is \emph{exact} by obeying equation \eqref{eq:EoM:H:Exact}.
\begin{align}
    \dd H\indices{_{\alpha\beta\gamma\delta}} = \varepsilon\indices{_{\alpha\beta\gamma\sigma}} K\indices{^{,\sigma}_\delta} -  \varepsilon\indices{_{\delta\alpha\beta\sigma}} K\indices{^{,\sigma}_\gamma} + \varepsilon\indices{_{\gamma\delta\alpha\sigma}} K\indices{^{,\sigma}_\beta} - \varepsilon\indices{_{\beta\gamma\delta\sigma}} K\indices{^{,\sigma}_\alpha} = 0.
\end{align}
Since $\varepsilon\indices{_{\alpha\beta\gamma\delta}}$ is fully antisymmetric, it follows that $\varepsilon\indices{_{\alpha\beta\gamma\delta}} \neq 0$ iff $\alpha\neq\beta\neq\gamma\neq\delta$. Hence, expressions of the form $\varepsilon\indices{_{\alpha\beta\gamma\sigma}} K\indices{^{,\sigma}_\delta}$ only get contributions when $\sigma=\delta$, all other terms cancelling, and giving all the terms a common factor of $K\indices{^{,\sigma}_\sigma}$.\footnote{This factor is \emph{not} summed over $\sigma$.}
\begin{align}
    \dd H\indices{_{\alpha\beta\gamma\delta}} & = \varepsilon\indices{_{\alpha\beta\gamma\delta}} K\indices{^{,\delta}_\delta} - \varepsilon\indices{_{\delta\alpha\beta\gamma}} K\indices{^{,\gamma}_\gamma} + \varepsilon\indices{_{\gamma\delta\alpha\beta}} K\indices{^{,\beta}_\beta} - \varepsilon\indices{_{\beta\gamma\delta\alpha}} K\indices{^{,\alpha}_\alpha} \nonumber \\
    & = \varepsilon\indices{_{\alpha\beta\gamma\delta}} \left( K\indices{^{,\delta}_\delta} + K\indices{^{,\gamma}_\gamma} + K\indices{^{,\beta}_\beta} + K\indices{^{,\alpha}_\alpha} \right) = \varepsilon\indices{_{\alpha\beta\gamma\delta}} \dalembertian K = 0.
\end{align}
Thus the solution for the $H$-field is described by the gradient of a massless scalar field
\begin{align}
    \begin{split}
    H\indices{_{\alpha\beta\gamma}} & = \varepsilon\indices{_{\alpha\beta\gamma\delta}} K\indices{^{,\delta}},
    \\ \quad \qq*{where} K\indices{_{,\mu}^{\mu}} & = 0 .\end{split} \label{eq:EoM:Lambda}
\end{align}
It also follows that the action can be rewritten to that of a massless scalar field\footnote{To obtain this result one needs the relation $\varepsilon\indices{^{\mu\nu\rho}_{\sigma}} \varepsilon\indices{_{\mu\nu\rho\delta}} = -6 g\ind{_{\sigma\delta}}$. This relation can be argued for from $\varepsilon\indices{^{\mu\nu\rho\sigma}} \varepsilon\indices{_{\mu\nu\rho\sigma}}=-4!$ in 4 space-time dimensions. Then, it follows straightforwardly that $\varepsilon\indices{^{\mu\nu\rho\sigma}} \varepsilon\indices{_{\mu\nu\rho\sigma}}=g\indices{^{\sigma\delta}} \left(\varepsilon\indices{^{\mu\nu\rho}_{\sigma}} \varepsilon\indices{_{\mu\nu\rho\delta}}\right) = -24 = g\indices{^{\sigma\delta}} \left(-6 g\ind{_{\sigma\delta}} \right)$. }
\begin{align}
    \begin{split}
        H\indices{_{\mu\nu\rho}} H\indices{^{\mu\nu\rho}} & = g\ind{^{\sigma\alpha}} \varepsilon\indices{_{\mu\nu\rho\sigma}} \varepsilon\indices{^{\mu\nu\rho}_{\delta}} g\ind{^{\delta\beta}} K\indices{_{,\alpha}} K\indices{_{,\beta}} = -6 g\ind{^{\sigma\alpha}} g\indices{_{\sigma\delta}} g\ind{^{\delta\beta}} K\indices{_{,\alpha}} K\indices{_{,\beta}} = -6 K\indices{_{,\sigma}} K\indices{^{,\sigma}}, \\
        \rightarrow \quad cS_B & = \frac{1}{12} \int \dd[4]{x} \sqrt{-\abs{g}} H\indices{_{\mu\nu\rho}} H\indices{^{\mu\nu\rho}} = -\frac{1}{2}\int \dd[4]{x} \sqrt{-\abs{g}} K\indices{_{,\sigma}} K\indices{^{,\sigma}}.
    \end{split} \label{eq:action:K}
\end{align}

\subsection{Relativistic expansion of the action}
Expressing the Kalb-Ramond action in terms of the scalar field $K$, like in equation \eqref{eq:action:K}, one can relativistically expand the action in the so-called \emph{post-Newtonian (PN)} expansion. In post-Newtonian calculations, the metric is split into the usual weak-field approximation $g\ind{_{\mu\nu}} = \eta\ind{_{\mu\nu}} + \lambda h\ind{_{\mu\nu}}$, where $ \eta\ind{_{\mu\nu}} $ is the Minkowski metric, choosing suitable Cartesian coordinates such that $\eta\ind{_{\mu\nu}} = \text{diag}(-1,1,1,1)$. $h\ind{_{\mu\nu}}$ denotes the metric perturbations, and will also be referred to here as the graviton field. $\lambda$ is a scaling factor that governs the relativistic expansion, which will be shown to be $\propto \sqrt{G}$, the Newtonian gravitational constant.

Using \eqref{eq:action:K}, the metric splitting gives an expansion of the Kalb-Ramond action in terms of the weak-(gravitational)field as
\begin{subequations} \label{eq:action:K:expanded}
\begin{align}
    cS_K & = -\frac{1}{2} \int \dd[4]{x} \sqrt{-\abs{g}} \left( \eta\indices{^{\mu\nu}} + \lambda h\indices{^{\mu\nu}} \right) K\indices{_{,\mu}} K\indices{_{,\nu}} \\
    & = -\frac{1}{2} \int \dd[4]{x} \left( 1 + \frac{1}{2} \lambda h +\dots \right) \left( \eta\indices{^{\mu\nu}} + \lambda h\indices{^{\mu\nu}} \right) K\indices{_{,\mu}} K\indices{_{,\nu}} \\
    & \approx -\frac{1}{2} \int \dd[4]{x} \left[ \left(1 + \frac{1}{2}\lambda h\right) K\indices{_{,\mu}} K\indices{^{,\mu}} + \lambda h\indices{^{\mu\nu}} K\indices{_{,\mu}} K\indices{_{,\nu}} \right]
\end{align}
\end{subequations}

The determinant of the metric was expanded using the following steps, assuming $\lambda h \ll 1$ and using $\eta\indices{_{\mu\nu}}$ for raising and lowering indices, which will be the case for the rest of this paper.
\begin{align}
\begin{split}
	&\det(g\indices{_{\rho\nu}}) = \det( \eta\indices{_{\rho\mu}} g\indices{^\mu_{\nu}}) =  \det(\eta\indices{_{\rho\mu}}) \det(g\indices{^\mu_{\nu}}) = -\det(g\indices{^\mu_{\nu}}) = -\det(\delta\indices{^\mu_\nu} + \lambda h\indices{^\mu_\nu} )\\
	& = -\exp(\ln( \det(\delta\indices{^\mu_\nu} + \lambda h\indices{^\mu_\nu} ) )) = -\exp(\tr( \ln(\delta\indices{^\mu_\nu} + \lambda h\indices{^\mu_\nu} ) )) \approx -\exp(\tr( \lambda h\indices{^\mu_\nu} )) \approx -1 - \lambda h.
\end{split} \label{eq:det(g)}
\end{align}
Then by also Taylor expanding the square root the following relation is also established, which appears in the action integral.
\begin{align}
	\sqrt{-\det(g)} \approx \sqrt{1+\lambda h} \approx 1 + \frac{\lambda}{2} h + \order{\lambda^2 h^2}. \label{eq:sqrt(det(g))}
\end{align}

The form of the action \eqref{eq:action:K:expanded} is fairly similar to that which is investigated in \cite[eq. (18)]{AxionLIGO} and \cite[eq. (3.18)]{AxionMaster} for massive axions coupling to gravity. The main difference is the absence of a mass term for the Kalb-Ramond field.

For the coupling term to sources like fuzzballs, we introduce effective Kalb-Ramond charges $q_a$, $p_a$, etc and generic interaction terms:
\begin{align}
    S_{\text{int}K} = -\sum_{a} \int \frac{\dd{c\tau_a}}{c^2} \dd[3]{x} \left[ q_a K + p_a K^2 +\dots \right] \sqrt{ -g\ind{_{\mu\nu}} \dot{x}^\mu \dot{x}^\nu } \dirac{3}{\vb*{x}-\vb*{x}_a(\tau_a)}. \label{eq:action:interaction:LambdaPP}
\end{align}
The term in the square root is needed to make the point particle action reparametrization invariant.\footnote{Which is required by relativity since there are no privileged time coordinates, and to have classical particles have the correct number of degrees of freedom. This requirement is also equivalent to the condition $\dot{x}\ind{^\alpha} \pdv{L}{\dot{x}\ind{^\alpha}} = L$ \cite[pp. $350-352$]{SpesRel}.} 

\subsection{The graviton and point-particle action}
The linearized action of gravity is \cite{Feynman:GravityLectures,ThirdOrder}
\begin{align}
	cS_h = \int \dd[4]{x} \left( -\frac{1}{4} h\indices{_{\mu\nu,\rho}}h\indices{^{\mu\nu,\rho}} + \frac{1}{8} h\indices{_{,\mu}} h\indices{^{,\mu}} + \order{h^3}\right), \label{eq:Graviton:Action}
\end{align}
with the effective coupling to matter via its energy-momentum tensor $T\indices{^{\mu\nu}}$:
\begin{align}
	cS_{\text{int}h} = \int \dd[4]{x} \frac{\lambda}{2} h\indices{_{\mu\nu}} T\indices{^{\mu\nu}}.
\end{align}
For objects influenced only by gravity, the action can be approximated at large scales by the free point-particle action from GR:
\begin{align}
	& S_{fpp} = -\sum_a m_a c \int \dd{s_a} = -\sum_{a} \int m_a c \sqrt{-g\indices{_{\mu\nu}} \dd{x}^\mu_a \dd{x}^\nu_a } \nonumber \\ & \hspace{0.7cm} = -\sum_{a} m_a c \int \sqrt{1-\lambda h\indices{_{\mu\nu}} \frac{ \dot{x}_a^\mu(x^0) }{c} \frac{ \dot{x}_a^\nu(x^0) }{c}}  \gamma_a^{-1} \dd{x^0} \nonumber \\
	& = \sum_{a} \int \frac{\dd[4]{x}}{c} \gamma_a^{-1} \left[ -m_a c^2 + \frac{\lambda}{2} h\indices{_{\mu\nu}} m_a \dot{x}\indices{^\mu} \dot{x}\indices{^\nu} + \frac{m_a\lambda^2}{8c^2} \left( h\indices{_{\mu\nu}} \dot{x}\indices{^\mu} \dot{x}\indices{^\nu} \right)^2 +\dots \right] \dirac{3}{ \vb*{x} - \vb*{x}_a(x^0) } .
\end{align}
The point-particle action can be further extended to include the effective coupling to the $K$-field most simply by adding the interaction term \eqref{eq:action:interaction:LambdaPP}
\begin{align}
\begin{split}
    S_{pp} = S_{fpp} + S_{\text{int}K} = \sum_a & \int \frac{\dd[4]{x}}{c} \gamma_a^{-1} \left(m_a c^2 + q_a K + p_a K^2 +\dots \right) \\
    & \cdot \left[ -1 + \frac{1}{2} \lambda h\indices{_{\mu\nu}} \frac{\dot{x}\indices{^\mu}}{c} \frac{\dot{x}\indices{^\nu}}{c} + \frac{1}{8} \left( \lambda h\indices{_{\mu\nu}} \frac{\dot{x}\indices{^\mu}}{c} \frac{\dot{x}\indices{^\nu}}{c} \right)^2 +\dots \right] \dirac{3}{ \vb*{x} - \vb*{x}_a(x^0) }.
\end{split} \label{eq:ppAction full}
\end{align}
The different cross-terms in this action will here be represented by Feynman diagrams in figure \ref{fig:1PN:H-type diagrams}-\ref{fig:3Point:diagrams}. Note that for multiple massive point particles the energy-momentum tensor is $T\ind{^{\mu\nu}}(x) = \sum_a \gamma_a^{-1} m_a \dot{x}\ind{^\mu} \dot{x}\ind{^\nu} \dirac{3}{\vb*{x} -\vb*{x}_a } $, which appears in the cross-term between the mass term and the linear term in $\lambda h\ind{_{\mu\nu}}$, hence the interaction term of gravity is to linear order $\frac{\lambda}{2} h\ind{_{\mu\nu}} T\ind{^{\mu\nu}}$.

\subsection{Equations of motion for the gravity-Kalb-Ramond sector}
Let $S_\text{total} = S_K + S_h + S_{pp} = \int \mathcal{L} \dd[4]{x}/c$. Variation of this action produces the equations of motion,
\begin{align}
    & \partial_\mu \fdv{\mathcal{L}}{h\indices{_{\alpha\beta,\mu}}} = -\frac{1}{2} \dalembertian h\indices{^{\alpha\beta}} + \frac{1}{4}\eta\ind{^{\alpha\beta}} \dalembertian h = \fdv{\mathcal{L}}{h\indices{_{\alpha\beta}}} = \frac{\lambda}{2}\left(T\indices{^{\alpha\beta}} + t^{\alpha\beta}\right) - \frac{\lambda}{2} \left( \frac{1}{2} \eta\indices{^{\alpha\beta}} K_{,\sigma} K^{,\sigma} + K^{,\alpha} K^{,\beta} \right),\nonumber \\ 
    & \Rightarrow \quad \dalembertian h\indices{_{\mu\nu}} = -\lambda P\indices{_{\mu\nu\alpha\beta}} \left( T\indices{^{\alpha\beta}} + t\indices{^{\alpha\beta}} - \frac{1}{2} \eta\indices{^{\alpha\beta}} K_{,\sigma} K^{,\sigma} - K^{,\alpha} K^{,\beta} \right) \label{eq:EoM:h} \\
    \mathrm{and} \nonumber \\
    &\partial_\mu \fdv{\mathcal{L}}{K\indices{_{,\mu}} } = -\partial_\mu  \left[ \left(1+\frac{\lambda}{2}h\right) K\indices{^{,\mu}} + \lambda h\indices{^{\mu\sigma}} K\indices{_{,\sigma}} \right] = \fdv{\mathcal{L}}{K} = -\overbrace{\sum_{a} \gamma_a^{-1} \left[ q_a + 2p_a K \right] \dirac{3}{\vb*{x} - \vb*{x}_a(x^0)} }^{=J_K} \nonumber \\
    &\Rightarrow \quad \dalembertian K = J_K - \lambda \left( h\indices{^{\mu\nu}} K\indices{_{,\mu\nu}} + h\indices{^{\mu\nu}_{,\mu}} K\indices{_{,\nu}} \right) - \frac{\lambda}{2} \left(h\indices{_{,\mu}} K^{,\mu} + h\dalembertian K \right) \label{eq:EoM:K}
\end{align}
The $P\indices{_{\mu\nu\alpha\beta}}$ is the combination of $\eta\ind{_{\mu\nu}}$'s which when acted upon the first equality of equation \eqref{eq:EoM:h} simplifies it to just $-\frac{1}{2}\dalembertian h\ind{_{\mu\nu}}$.\footnote{For those familiar with the \emph{bar operator} introduced in \cite{Feynman:GravityLectures}, $P\indices{_{\mu\nu}^{\alpha\beta}}$ is the bar operator, which is also its own inverse.} The exact expression for $P\indices{_{\mu\nu\alpha\beta}}$ can be found in equation \eqref{eq:Projection tensor}. $t\indices{^{\alpha\beta}}$ represents the energy-momentum tensor of the graviton field itself, arising from cubic terms in the graviton action \eqref{eq:Graviton:Action}. To leading order $t\indices{_{\alpha}^{\beta}} = \frac{1}{2} h\ind{_{\mu\nu,\alpha}}h\ind{^{\mu\nu,\beta}}$.

Both of the equations \eqref{eq:EoM:h} and \eqref{eq:EoM:K} can be solved by the method of Green's functions.
\begin{align}
    & h\indices{_{\mu\nu}}(x) = -\lambda \int \dd[4]{y} \Delta(x^\sigma - y^\sigma) P\indices{_{\mu\nu\alpha\beta}} \left(T\indices{^{\alpha\beta}} + t\indices{^{\alpha\beta}} - \frac{1}{2} \eta\indices{^{\alpha\beta}} K_{,\sigma} K^{,\sigma} - K^{,\alpha} K^{,\beta} \right), \label{eq:FieldExpansion:h:general} \\
    & K(x) = \int \dd[4]{y} \Delta(x^\sigma-y^\sigma) \left(  J_K - \lambda \left( h\indices{^{\mu\nu}} K\indices{_{,\mu\nu}} + h\indices{^{\mu\nu}_{,\mu}} K\indices{_{,\nu}} \right) - \frac{\lambda}{2} \left(h\indices{_{,\mu}} K^{,\mu} + h\dalembertian K \right) \right), \label{eq:FieldExpansion:K:general}
\end{align}
where $\Delta(x-y)$ is the Green's function of the d'Alembertian: 
\begin{align}
    & \Delta(x^\sigma-y^\sigma) = \int \frac{\dd[4]{k}}{(2\pi)^4} \frac{-1}{ k\indices{_\sigma} k\indices{^\sigma} } e^{ik_\mu \left( x^\mu - y^\mu \right)}, \label{eq:GreensFunction:d'Alembert} \\
	& P\indices{_{\mu\nu\alpha\beta}} = \frac{1}{2}\left( \eta_{\mu\alpha} \eta_{\nu\beta} + \eta_{\mu\beta}\eta_{\nu\alpha} - \eta_{\mu\nu} \eta_{\alpha\beta} \right). \label{eq:Projection tensor}
\end{align}
To compute the leading order terms of the action in the post-Newtonian formalism interactions between different and similar fields can be neglected. Thus only the leading term of \eqref{eq:FieldExpansion:h:general} and \eqref{eq:FieldExpansion:K:general} contribute. Also to get the static, non-retarded, Newtonian potential the Green's function \eqref{eq:GreensFunction:d'Alembert} must be expanded in a power series and truncated at leading order. This effectively approximates it as the Green's function of the \emph{Laplacian} operator $\nabla^2=\partial\ind{_i}\partial\ind{^i}$ (with corrections scaling as $(v/c)^{2n}$):
\begin{align}
    & \Delta(x^\mu-y^\mu) = \int \frac{\dd[4]{k}}{(2\pi)^4} \frac{ e^{ik_\mu \left( x^\mu - y^\mu \right)} }{ -k_\sigma k^\sigma} = \int_k \frac{ e^{ik_\mu \left( x^\mu - y^\mu \right)} }{ k_0^2 - \vb*{k}^2 } = \int_k \frac{ e^{ik_\mu \left( x^\mu - y^\mu \right)} }{ -\vb*{k}^2 \left( 1 - \frac{k_0^2}{\vb*{k}^2} \right) } \nonumber \\
	&\approx \int_k \frac{ e^{ik_\mu \left( x^\mu - y^\mu \right)} }{ -\vb*{k}^2 } \left( 1 + \left(\frac{k_0}{\vb*{k}}\right)^2 +  \left(\frac{k_0}{\vb*{k}}\right)^4 + \dots \right) = \Delta^{(0)}_\text{inst}(x-y) + \Delta^{(2)}_\text{inst}(x-y) + \dots  \label{eq:Propagator expansion}\\
    & = \frac{-\delta (x^0 - y^0)}{4\pi \abs{\vb*{x} -\vb*{y}} } - \frac{ \vb*{\dot{x}} \vb*{\cdot} \vb*{\dot{y}} - (\vb*{\dot{x}}\vb*{\cdot} \vb*{\hat{r}})(\vb*{\dot{y}}\vb*{\cdot} \vb*{\hat{r}}) }{8\pi c^2 \abs{\vb*{x} -\vb*{y}} }\delta (x^0 - y^0) +\dots = \frac{-\delta(x^0 - y^0)}{4\pi r} \left( 1 - \frac{\eta}{2} \frac{\vb*{v}^2 - (\vb*{v\cdot \hat{r}})^2 }{ c^2 } + \dots \right). \nonumber
\end{align}
\begin{figure}
    \includegraphics[width=0.9\textwidth]{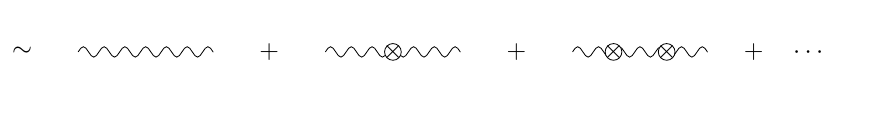}
    \caption{Diagrammatically the velocity expansion \eqref{eq:Propagator expansion} of the Green's function is represented by the $\otimes$-symbol on top of the propagator \cite{Porto2016}. The number of $\otimes$ on the line indicates which order in the expansion is being considered, the leading order Newtonian propagator having none, the 1PN having one, etc.}
\end{figure}
Here the superscript (0) in $\Delta^{(0)}_\text{inst}(x-y)$ is to indicate this term scale as $(v/c)^0$ and thus belong in the 0PN order, while $\Delta^{(2)}_\text{inst}(x-y)$ scale as $(v/c)^2$ and belongs to the 1PN. It is given the subscript `inst' as this approximation makes the interaction instantaneous (via the factor of $\delta (x^0 - y^0)$).

With this all in mind leading-order terms of the action governed by graviton and Kalb-Ramond fields are
\begin{subequations}
\begin{align}
    &S_{pp}^{(0)} = \int \frac{\dd[4]{x}}{c} \sum_{a} \gamma_a^{-1} \left[ -m_ac^2 + \frac{m_a \lambda}{2} h_{00}(x) \dot{x}^0 \dot{x}^0 - q_a K(x) \right] \dirac{3}{ \vb*{x} - \vb*{x}_a(x^0) } \\
    &= \int \frac{\dd[4]{x}}{c} \sum_{a} \gamma_a^{-1} \Biggl[ \sum_{b>a} \left\{ \frac{m_a \lambda}{2} \frac{\lambda m_b \gamma_b c^2}{8\pi \abs{\vb*{x} - \vb*{x}_b }} (\gamma_a c)^2 + q_a \frac{\gamma_b^{-1} q_b}{4\pi\abs{\vb*{x} - \vb*{x}_b }} \right\} -m_ac^2 \Biggr] \dirac{3}{ \vb*{x} - \vb*{x}_a(x^0) } \\
    &= \int \left[ \frac{1}{2}m_1 v_1^2 + \frac{1}{2}m_2 v_2^2 + \frac{\lambda^2 m_1c^2 m_2c^2}{16\pi r} + \frac{q_1 q_2 }{4\pi r} \right] \dd{t}.
\end{align}
\end{subequations}
Thus we recognise the action of point particles in a Newtonian gravitational potential, also interacting with a Coulomb-like potential. The Newtonian limit implies 
\begin{align}
	\lambda^2 = \frac{16\pi G}{c^4} = \frac{16\pi}{ F_\text{Pl}}, \qq{$F_\text{Pl}$ being ``the Planck force''.}
\end{align}

\section{Expanding the action into the 1PN order}
\subsection{Velocity corrections to the 0PN diagrams}
\begin{figure}[!ht]
    \begin{center}
        \begin{subfigure}[b]{0.22\textwidth}
            \centering
            \includegraphics[width=\textwidth]{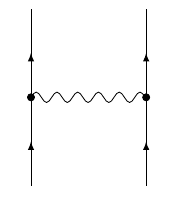}
            \caption{}
            \label{fig:Feynman:H-type:0PN}
        \end{subfigure}
        \begin{subfigure}[b]{0.35\textwidth}
            \centering
            \includegraphics[width=\textwidth]{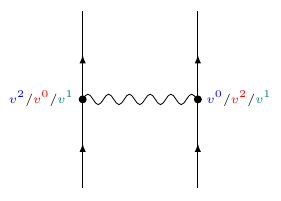}
            \caption{}
            \label{fig:Feynman:H-type:VelocityExp}
        \end{subfigure}
        \begin{subfigure}[b]{0.22\textwidth}
            \centering
            \includegraphics[width=\textwidth]{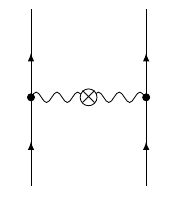}
            \caption{}
            \label{fig:Feynman:H-type:ox}
        \end{subfigure}
    \end{center}
    
    \begin{center}
        \begin{subfigure}[b]{0.22\textwidth}
            \centering
            \includegraphics[width=\textwidth]{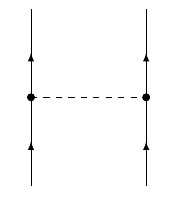}
            \caption{}
            \label{fig:Feynman:H-type:0PN:K}
        \end{subfigure}
        \begin{subfigure}[b]{0.35\textwidth}
            \centering
            \includegraphics[width=\textwidth]{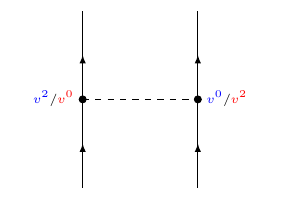}
            \caption{}
            \label{fig:Feynman:H-type:VelocityExp:K}
        \end{subfigure}
        \begin{subfigure}[b]{0.22\textwidth}
            \centering
            \includegraphics[width=\textwidth]{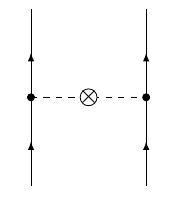}
            \caption{}
            \label{fig:Feynman:H-type:ox:K}
        \end{subfigure}
    \end{center}
        \caption{Feynman diagrams of the 0PN potential of GR (\subref{fig:Feynman:H-type:0PN}) and Kalb-Ramond (\subref{fig:Feynman:H-type:0PN:K}), with 1PN velocity corrections (\subref{fig:Feynman:H-type:VelocityExp})-(\subref{fig:Feynman:H-type:ox}) and (\subref{fig:Feynman:H-type:VelocityExp:K})-(\subref{fig:Feynman:H-type:ox:K}). The wavy lines represent graviton propagators, the dashed lines Kalb-Ramond propagators, and the solid lines represent non-propagating, point-particle, world-line sources \cite{Goldberger:EFT}.}
        \label{fig:1PN:H-type diagrams}
\end{figure}
In order to obtain the point particle action contribution at next to leading order in $(v/c)^{2n}$, the so-called 1PN order, both velocity modifications to the previous potentials as well as higher order terms of the potential must be accounted for. Diagrammatically the velocity corrections to the leading order term can be expressed using the Feynman diagrams of figure \ref{fig:1PN:H-type diagrams}. Figure \ref{fig:Feynman:H-type:0PN} and \ref{fig:Feynman:H-type:0PN:K} represent the 0PN contributions already calculated
\begin{subequations}
\begin{align}
    V_{\ref{fig:Feynman:H-type:0PN}} &= -\frac{Gm_1 m_2}{r}, \\
    V_{\ref{fig:Feynman:H-type:0PN:K}} &= -\frac{q_1 q_2}{4\pi r}.
\end{align}
\end{subequations}
Diagrams \ref{fig:Feynman:H-type:ox} and \ref{fig:Feynman:H-type:ox:K} belongs to the 1PN and represent corrections to the Green's function as described in equation \eqref{eq:Propagator expansion}:
\begin{subequations}
\begin{align}
    V_{\ref{fig:Feynman:H-type:ox}} &= -\frac{Gm_1 m_2}{r}\left( \frac{\vb*{v}_1\vb*{\cdot} \vb*{v}_2 - \left( \vb*{v}_1 \vb*{\cdot} \vb*{\hat{r}} \right) \left( \vb*{v}_2 \vb*{\cdot} \vb*{\hat{r}} \right) }{2c^2} \right), \\
    V_{\ref{fig:Feynman:H-type:ox:K}} &= -\frac{q_1 q_2}{4\pi r}\left( \frac{\vb*{v}_1\vb*{\cdot} \vb*{v}_2 - \left( \vb*{v}_1 \vb*{\cdot} \vb*{\hat{r}} \right) \left( \vb*{v}_2 \vb*{\cdot} \vb*{\hat{r}} \right) }{2c^2} \right).
\end{align}
\end{subequations}
The last two diagrams, \ref{fig:Feynman:H-type:VelocityExp} and \ref{fig:Feynman:H-type:VelocityExp:K}, represent general velocity expansion of the interaction term, mainly the expansion of Lorentz factors $\gamma_a^{-1} = 1 - \frac{v_a^2}{2c^2} - \frac{v^4_a}{8c^4} - \dots$, but for the graviton interaction one must also include spatial contributions of the source four-velocities, like $h\ind{_{ij}} \dot{x}\ind{^i} \dot{x}\ind{^j}$ and  $h\ind{_{0i}} \dot{x}\ind{^0} \dot{x}\ind{^i}$.
\begin{subequations}
\begin{align}
    V_{\ref{fig:Feynman:H-type:VelocityExp}} &= \begin{cases}
        \textcolor{blue}{-\frac{Gm_1 m_2}{r} \frac{3\vb*{v}_1^2}{2c^2} } \\ \textcolor{red}{-\frac{Gm_1 m_2}{r} \frac{3\vb*{v}_2^2}{2c^2} } \\ \textcolor{teal}{\phantom{-}\frac{Gm_1 m_2}{r} \frac{4\vb*{v}_1 \vb*{\cdot} \vb*{v}_2}{c^2} }
    \end{cases}, \\
    V_{\ref{fig:Feynman:H-type:VelocityExp:K}} &= \begin{cases}
        \textcolor{blue}{\phantom{-}\frac{q_1 q_2}{4\pi r} \frac{\vb*{v}_1^2}{2c^2} } \\ \textcolor{red}{\phantom{-}\frac{q_1 q_2}{4\pi r} \frac{\vb*{v}_2^2}{2c^2} }
    \end{cases}.
\end{align}
\end{subequations}
For higher PN orders these expansions are continued and mixed, for example including a mix of diagram \ref{fig:Feynman:H-type:VelocityExp} and \ref{fig:Feynman:H-type:ox}, and a diagrams with two $\otimes$, and interaction points of the type $v^2 - v^2$ and $v^4 - v^0$ in analogy to diagrams \ref{fig:Feynman:H-type:VelocityExp} and \ref{fig:Feynman:H-type:VelocityExp:K}. But for the 1PN, figure \ref{fig:1PN:H-type diagrams} contains all the contributing single propagator diagrams.

\subsection{Higher order diagrams: seagull}
\begin{figure}[!ht]
    \begin{center}
        \begin{subfigure}[b]{0.22\textwidth}
            \centering
            \includegraphics[width=\textwidth]{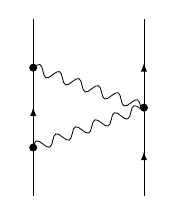}
            \caption{}
            \label{fig:Feynman:Seagull:HH}
        \end{subfigure}
        \begin{subfigure}[b]{0.22\textwidth}
            \centering
            \includegraphics[width=\textwidth]{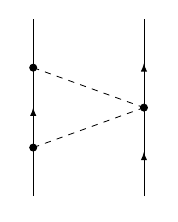}
            \caption{}
            \label{fig:Feynman:Seagull:KK}
        \end{subfigure}
        \begin{subfigure}[b]{0.22\textwidth}
            \centering
            \includegraphics[width=\textwidth]{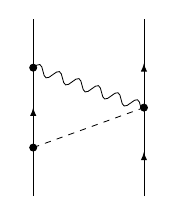}
            \caption{}
            \label{fig:Feynman:Seagull:HK}
        \end{subfigure}
    \end{center}
        \caption{\emph{Seagull}-type diagrams contributing at 1PN. These diagrams originate from higher-order terms in the interaction expansion of the point particle action, see equation \eqref{eq:ppAction full}.}
        \label{fig:Seagull diagrams}
\end{figure}
Moving on to higher order diagrams, that is diagrams of higher order powers of the fields $h\ind{_{\mu\nu}}$ and $K$. Figure \ref{fig:Seagull diagrams} depicts the next-to-leading order interaction terms of equation \eqref{eq:ppAction full}, namely $\text{fig. } \ref{fig:Feynman:Seagull:HH} \simeq m_a c^2 \times \frac{1}{8} \left( \lambda h\indices{_{\mu\nu}} \frac{\dot{x}\indices{^\mu}}{c} \frac{\dot{x}\indices{^\nu}}{c} \right)^2$, $\text{fig. } \ref{fig:Feynman:Seagull:KK} \simeq p_a K^2 \times \left(-1\right)$, and $\text{fig. } \ref{fig:Feynman:Seagull:HK} \simeq q_a K \times \frac{1}{2}\lambda h\indices{_{\mu\nu}} \frac{\dot{x}\indices{^\mu}}{c} \frac{\dot{x}\indices{^\nu}}{c}$. Inserting the leading order solution for $\lambda h\indices{_{00}}(x) = \sum_b \frac{\lambda^2 m_b \gamma_b c^2 \delta(x^0 - x^0_b)}{8\pi \abs{\vb*{x}-\vb*{x}_b }}$ and $K(x) = \sum_b -\frac{\gamma_b^{-1} q_b \delta(x^0 - x^0_b)}{4\pi \abs{\vb*{x} - \vb*{x}_b}}$, the resulting terms of the action originating from these diagrams and subsequently their potential is easily obtained.
\begin{subequations}
\begin{align}
    V_{\ref{fig:Feynman:Seagull:HH}} &= -\frac{ G^2 (m_1 m_2^2 + m_2 m_1^2) }{ 2 c^2 r^2 }, \\
    V_{\ref{fig:Feynman:Seagull:KK}} &= \phantom{-} \frac{ (p_1 q_2^2 + p_2 q_1^2) }{ 16 \pi^2 r^2 }, \\
    V_{\ref{fig:Feynman:Seagull:HK}} &= \phantom{-} \frac{ Gq_1 q_2 (m_1+m_2) }{ 4\pi c^2 r^2}.
\end{align}
\end{subequations}

\subsection{Higher order diagrams: field-interactions}
\begin{figure}[!ht]
    \begin{center}
        \begin{subfigure}[b]{0.22\textwidth}
            \centering
            \includegraphics[width=\textwidth]{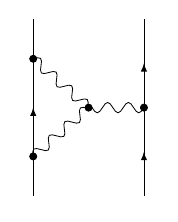}
            \caption{}
            \label{fig:Feynman:3p:HHH}
        \end{subfigure}
        \begin{subfigure}[b]{0.22\textwidth}
            \centering
            \includegraphics[width=\textwidth]{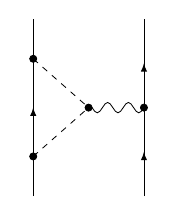}
            \caption{}
            \label{fig:Feynman:3p:KKH}
        \end{subfigure}
        \begin{subfigure}[b]{0.22\textwidth}
            \centering
            \includegraphics[width=\textwidth]{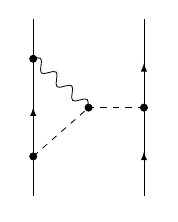}
            \caption{}
            \label{fig:Feynman:3p:HKK}
        \end{subfigure}
    \end{center}
        \caption{The diagrams with \emph{field-interactions} contributing at 1PN. Wavy lines denote graviton propagators and dashed lines denote propagators of the $K$-scalar Kalb-Ramond field.}
        \label{fig:3Point:diagrams}
\end{figure}
The last set of diagrams contributing to the 1PN action originates from corrections to the solution of the fields themselves, see equation \eqref{eq:FieldExpansion:h:general}-\eqref{eq:FieldExpansion:K:general}, and are depicted diagrammatically in figure \ref{fig:3Point:diagrams} as \emph{field-interactions}. The explicit form of the action term is here
\begin{align}
	S_{\text{fig.}\ref{fig:3Point:diagrams}} = \int \frac{\dd[4]{x}}{c} \sum_{a} \gamma_a^{-1} \left[ \frac{m_a \lambda}{2}h_{00}^{(2)}(x) (\gamma_a c)^2 - q_a K^{(2)}(x) \right] \dirac{3}{\vb*{x} - \vb*{x}_a(x^0)}, \label{eq:ThreePoint:Action:intro}
\end{align}
where the corrections in the fields $h_{00}^{(2)}(x)$ and $K^{(2)}(x)$ are the items of interest.

\subsubsection{The corresponding potential of diagram \ref{fig:Feynman:3p:HHH}}
Diagram \ref{fig:Feynman:3p:HHH} originates from the graviton energy-momentum (pseudo-)tensor $t\ind{^{\alpha\beta}}$ contribution to the equation of motion \eqref{eq:EoM:h} with the corresponding solution:
\begin{align}
\begin{split}
    h^{(2)}_{00}(x) \ni -\lambda \int \dd[4]{y} \Delta(x^\sigma - y^\sigma) P\indices{_{00\alpha\beta}} \frac{1}{2} h\ind{_{\mu\nu}^{,\alpha}} h\ind{^{\mu\nu,\beta}}.
\end{split}
\end{align}
To solve this, one can write out the $h\ind{_{\mu\nu}}$ fields as their leading order solution in Fourier space 
\begin{align}
    h\ind{_{\mu\nu}^{,\alpha}}(y) = & -\lambda \iint \frac{\dd[4]{p}}{(2\pi)^4} \frac{ ip\ind{^\alpha} e^{i k\ind{_{\rho}} \left( y\ind{^{\rho}} - z\ind{^{\rho}} \right) }}{ -p\ind{_{\sigma}} p\ind{^{\sigma}} } P\ind{_{\mu\nu00}} T\ind{^{00}}(z) \dd[4]{z}, \\
    & P\ind{_{\mu\nu00}} = P\ind{_{00\mu\nu}} = \frac{1}{2} \delta\ind{_{\mu\nu}}.
\end{align}
For algebraic convenience, one might also utilise the following relation valid for any field $\Phi$ and $\Theta$:
\begin{align}
    \Phi\ind{_{,\{\alpha}} \Theta\ind{_{,\beta\}}} = \frac{1}{2}\left( \left( 
\Phi \Theta \right)\ind{_{,\alpha\beta}} - \Phi\ind{_{,\alpha\beta}} \Theta - \Phi \Theta\ind{_{,\alpha\beta}} \right). \label{eq:derivative trick}
\end{align}
Using the notation $\int \frac{\dd[4]{k}}{(2\pi)^4} = \int_k$, and taking advantage of rewriting the derivatives in accordance with \eqref{eq:derivative trick} one obtains
\begin{align}
\begin{split}
    h^{(2)}_{00}(x) \ni -\lambda^3 \int \int_k \int_{p_1, p_2} \frac{ e^{ik_\sigma(x^\sigma - y^\sigma) } }{ k\ind{_\sigma} k\ind{^\sigma} } \frac{ e^{ip_{1\sigma}(y^\sigma - z_1^\sigma) } }{ p\ind{_{1\sigma}} p^{\sigma}_1 } \frac{ e^{ip_{2\sigma}(y^\sigma - z_2^\sigma) } }{ p\ind{_{2\sigma}} p^{\sigma}_2 } \frac{1}{8} \delta\ind{_{\alpha\beta}}\times \\
    \Biggl[ (p_1^\alpha+p_2^\alpha)(p_1^\beta+p_2^\beta)
    - p_1^\alpha p_1^\beta - p_2^\alpha p_2^\beta \Biggr] T\ind{^{00}}(z_1) T\ind{^{00}}(z_1) \dd[4]{y}
\end{split}
\end{align}
Separating out all the factors of, and integrating over $y$ results in a conservation of momentum factor
\begin{align*}
    \int e^{-iy\ind{^{\mu}}\left( k\ind{_\mu} - p_{1\mu} - p_{2\mu} \right) } \dd[4]{y} = (2\pi)^4 \delta\hspace{-2pt}\left(k\ind{_\mu} - p_{1\mu} - p_{2\mu} \right).
\end{align*}
Then performing the $k$ integral makes $k\ind{_\mu} = p_{1\mu} + p_{2\mu} \equiv q\ind{_\mu}$, and tidying up the expression gives
\begin{align}
\begin{split}
    h^{(2)}_{00}(x) \ni -\lambda^3 \int_{p_1, p_2} \frac{ e^{iq_\sigma x^\sigma } }{ q\ind{_\sigma} q\ind{^\sigma} } \frac{ e^{-ip_{1\sigma} z_1^\sigma } }{ p\ind{_{1\sigma}} p^{\sigma}_1 } \frac{ e^{-ip_{2\sigma} z_2^\sigma } }{ p\ind{_{2\sigma}} p^{\sigma}_2 } \frac{1}{8} \delta\ind{_{\alpha\beta}} \Biggl[ q^\alpha q^\beta - p_1^\alpha p_1^\beta - p_2^\alpha p_2^\beta \Biggr] T\ind{^{00}}(z_1) T\ind{^{00}}(z_2).
\end{split}
\end{align}
The final step is to again let $\frac{1}{k_\mu k^\mu} \to \frac{1}{\vb*{k}^2}$, and likewise ignore factors of $p^0$ appearing in the numerator as these also will result in velocity corrections and thus belong in higher PN order terms, see equation \eqref{eq:Propagator expansion}. Then we finally have
\begin{align}
    h^{(2)}_{00}(x) \ni -\frac{\lambda^3}{8} \int_{p_1, p_2} &\frac{ e^{iq_\sigma x^\sigma } }{ \vb*{q}^2 } \frac{ e^{-ip_{1\sigma} z_1^\sigma } }{ \vb*{p}_1^2 } \frac{ e^{-ip_{2\sigma} z_2^\sigma } }{ \vb*{p}_2^2 } \Biggl[ \vb*{q}^2 - \vb*{p}_1^2 - \vb*{p}_2^2 \Biggr] T\ind{^{00}}(z_1) T\ind{^{00}}(z_2) \nonumber \\
    = -\frac{\lambda^3}{8} \int_{p_1,p_2} &\Biggl[ \frac{ e^{ip_{1\sigma} (x^\sigma - z_1^\sigma) } }{ \vb*{p}_1^2 } \frac{ e^{ip_{2\sigma} (x^\sigma - z_2^\sigma) } }{ \vb*{p}_2^2 } - \frac{ e^{iq_{\sigma} (x^\sigma - z_1^\sigma) } }{ \vb*{q}^2 } \frac{ e^{ip_{2\sigma} (z_1^\sigma - z_2^\sigma) } }{ \vb*{p}_2^2 } \label{eq:htripp:intermediate}  \\
    &\quad \quad - \frac{ e^{iq_{\sigma} (x^\sigma - z_2^\sigma) } }{ \vb*{q}^2 } \frac{ e^{ip_{2\sigma} (z_2^\sigma - z_1^\sigma) } }{ \vb*{p}_1^2 } \Biggr] T\ind{^{00}}(z_1) T\ind{^{00}}(z_2). \nonumber
\end{align}
In the first term $\vb*{q}^2$, the first denominator is cancelled, and remembering that $q\ind{^\mu} = p_1^\mu + p_2^\mu$ the remaining exponential factor can be split up and distributed into the remaining exponential factors. Thus the first term looks like
\begin{align}
\begin{split}
    & -\frac{\lambda^3}{8} \int_{p_1, p_2} \frac{ e^{ip_{1\sigma} \left( x\ind{^\sigma} - z_1^\sigma \right) } }{ \vb*{p}_1^2 } \frac{ e^{ip_{2\sigma}\left( x\ind{^\sigma} - z_2^\sigma \right) } }{ \vb*{p}_2^2 }T\ind{^{00}}(z_1) T\ind{^{00}}(z_2) \\
    &= -\frac{\lambda^3}{8} \Delta_\text{inst}^{(0)}(x\ind{^\mu}-z_1^\mu) \Delta_\text{inst}^{(0)}(x\ind{^\mu}-z_2^\mu) T\ind{^{00}}(z_1) T\ind{^{00}}(z_2).
\end{split}
\end{align}
The procedure is the same for the two last terms of \eqref{eq:htripp:intermediate}, with the addition of shifting one of the integration variables to $q^\mu$.

Returning to the action \eqref{eq:ThreePoint:Action:intro} for diagram \ref{fig:Feynman:3p:HHH} it is
\begin{align}
    S_\text{fig.\ref{fig:Feynman:3p:HHH}} = \int \sum_{a\neq b,c} \frac{\lambda^4 m_a m_b m_c c^6}{16 (4\pi)^2} & \left( \frac{-1}{\abs{\vb*{x}_a - \vb*{z}_b} \cdot \abs{\vb*{x}_a - \vb*{z}_c}} + \frac{1}{\abs{\vb*{x}_a - \vb*{z}_b} \cdot \abs{\vb*{z}_b - \vb*{z}_c}} + \frac{1}{\abs{\vb*{x}_a - \vb*{z}_c} \cdot \abs{\vb*{z}_c - \vb*{z}_b}} \right) \dd{t} \nonumber \\
    &= \int -\frac{G^2 m_1 m_2(m_1+m_2)}{c^2 r^2} \dd{t}, \\
    V_{\ref{fig:Feynman:3p:HHH}} &= \frac{G^2 m_1 m_2(m_1 + m_2)}{c^2 r^2}. \label{eq:potential:3p:HHH}
\end{align}
The requirement $b\neq a \neq c$ of the sum is to make sure the gravitational potential acting on particle number $a$ is not sourced by itself. Discarding then all divergent terms\footnote{For more rigour impose particle coordinates before performing the momentum integrals. Then one obtains terms like $\int_{p_2} e^{ip_{2\sigma} ( 0^\sigma )}/\vb*{p}_2^2 = \int_{p_2} 1/\vb*{p}_2^2 $ which also diverge to infinity.} $\propto \frac{1}{\abs{\vb*{z}_i - \vb*{z}_i }}$ one is left with the potential \eqref{eq:potential:3p:HHH}.

\subsubsection{The corresponding potential of diagram \ref{fig:Feynman:3p:KKH}}
Diagram \ref{fig:Feynman:3p:KKH} is quite similar to diagram \ref{fig:Feynman:3p:HHH}, but rather accounts for the Kalb-Ramond scalar contribution to $h_{00}^{(2)}(x)$ from the equation of motion \eqref{eq:EoM:h}:
\begin{subequations}\label{eq:ThreePointGrav}
\begin{align}
	h^{(2)}_{00}(x) \ni & \lambda \int \dd[4]{y} \Delta(x^\sigma - y^\sigma) P\indices{_{00\alpha\beta}} \left\{ \frac{1}{2} \eta\indices{^{\alpha\beta}} K_{,\sigma}(y) K^{,\sigma}(y) + K^{,\alpha}(y) K^{,\beta}(y) \right\} \\
	\begin{split}
	& = \frac{i^2\lambda}{2} \int_{p_1, p_2} \frac{e^{iq_\mu x^\mu } }{-\vb*{q}^2 } \frac{e^{-ip_{1\mu} z_1^\mu } }{ -\vb*{p}_{1}^2 } \frac{e^{-ip_{2\mu} z_2^\mu } }{-\vb*{p}_{2}^2 } \Biggl[ \vb*{q}^2 - \vb*{p}_{1}^2 - \vb*{p}_2^2 \Biggr] J(z_1) J(z_2), 
	\end{split}
\end{align}
\end{subequations}
where again $q^\mu = p_1^\mu + p_2^\mu$. These integrals are identical to the ones found in \eqref{eq:htripp:intermediate}, and are solved the same way.
\begin{align}
	h_{00}^{(2)} \ni & \frac{\lambda}{2} \Bigl[ \Delta_\text{inst}(x-z_1) \Delta_\text{inst}(x-z_2) - \Delta_\text{inst}(x-z_1) \Delta_\text{inst}(z_1-z_2) - \Delta_\text{inst}(x-z_2) \Delta_\text{inst}(z_2-z_1)\Bigr] J(z_1) J(z_2) \nonumber \\
	& = \sum_{b, c} \frac{\lambda q_b q_c}{32 \pi^2}\left[ \frac{ 1 }{ \abs{ \vb*{x} - \vb*{z}_b} \cdot \abs{ \vb*{x} - \vb*{z}_c } } - \frac{ 1 }{ \abs{ \vb*{x} - \vb*{z}_b } \cdot \abs{ \vb*{z}_b - \vb*{z}_c } } - \frac{ 1 }{ \abs{ \vb*{x} - \vb*{z}_c } \cdot \abs{ \vb*{z}_c - \vb*{z}_b } } \right].
\end{align}
By substituting this result back into the action \eqref{eq:ThreePoint:Action:intro}, the associated potential can be read off.
\begin{align}
	S_\text{fig.\ref{fig:Feynman:3p:KKH}} = \int \sum_{a\neq b,c} \frac{ \lambda^2 m_a q_b q_c c^2}{64\pi^2} \Bigl( & \frac{ 1 }{ \abs{ \vb*{x} - \vb*{z}_b} \cdot \abs{ \vb*{x} - \vb*{z}_c } } - \frac{ 1 }{ \abs{ \vb*{x} - \vb*{z}_b } \cdot \abs{ \vb*{z}_b - \vb*{z}_c } } - \frac{ 1 }{ \abs{ \vb*{x} - \vb*{z}_c } \cdot \abs{ \vb*{z}_c - \vb*{z}_b } }\Bigr) \dd{t} \nonumber \\
	& = \int \frac{G(m_1q_2^2+m_2 q_1^2) }{4\pi r^2 c^2} \dd{t}, \\
    & V_{\ref{fig:Feynman:3p:KKH}} = - \frac{G(m_1q_2^2+m_2 q_1^2) }{4\pi r^2 c^2}.
\end{align}

\subsubsection{The corresponding potential of diagram \ref{fig:Feynman:3p:HKK}}
For diagram \ref{fig:Feynman:3p:HKK} it is the next-to-leading-order terms in the equation of motion for the Kalb-Ramond scalar \eqref{eq:EoM:K} which is included.
\begin{align}
	K^{(2)}(x) = -\lambda \int \dd[4]{y} \Delta(x^\sigma-y^\sigma) \left( h\indices{^{\mu\nu}} K\indices{_{,\mu\nu}} + h\indices{^{\mu\nu}_{,\mu}} K\indices{_{,\nu}}  + \frac{1}{2} \left(h\indices{_{,\mu}} K^{,\mu} + h\dalembertian K \right) \right). \label{eq:Lambda(2):correction:intro}
\end{align}

Since the derivative indices are either contracted with themselves or with the graviton field  $h^{\mu\nu}\partial_\mu \partial_\nu$ they are symmetric under $\mu \leftrightarrow \nu$, and one may therefore use \eqref{eq:derivative trick} also in this context. Using this and the other tricks used on diagrams \ref{fig:Feynman:3p:HHH} and \subref{fig:Feynman:3p:KKH} one can demonstrate
\begin{align}
\begin{split}
	K^{(2)}(x) = & -\frac{i^2\lambda^2}{2} \int_{p,k}  \frac{e^{iq_\sigma x^\sigma }}{-\vb*{q}^2} \frac{e^{-ik_\sigma z^\sigma_b }}{\vb*{k}^2} \frac{e^{-ip_\sigma z^\sigma_c }}{-\vb*{p}^2} \left( \vb*{q}^2 + \vb*{p}^2 - \vb*{k}^2 \right) T\ind{_{00}}(z_b) J(z_c) \label{eq:Lambda(2):correction} \\
    = & \sum_{b,c} \frac{\lambda^2 m_b q_c c^2}{32 \pi^2} \left( \frac{1}{\abs{ \vb*{x} - \vb*{z}_b } \cdot \abs{ \vb*{x} - \vb*{z}_c }} + \frac{1}{\abs{ \vb*{x} - \vb*{z}_c } \cdot \abs{ \vb*{z}_c - \vb*{z}_b }} - \frac{1}{\abs{ \vb*{x} - \vb*{z}_b } \cdot \abs{ \vb*{z}_b - \vb*{z}_c }}\right)
\end{split}
\end{align}
Again $q^\mu = k^\mu + p^\mu$.
\begin{align}
	S_\text{fig.\ref{fig:Feynman:3p:HKK}} = \int \sum_{a\neq b,c} \frac{ G q_a m_b q_c}{2\pi c^2} &\left( \frac{1}{\abs{ \vb*{x} - \vb*{z}_b } \cdot \abs{ \vb*{x} - \vb*{z}_c }} + \frac{1}{\abs{ \vb*{x} - \vb*{z}_c } \cdot \abs{ \vb*{z}_c - \vb*{z}_b }} - \frac{1}{\abs{ \vb*{x} - \vb*{z}_b } \cdot \abs{ \vb*{z}_b - \vb*{z}_c }}\right) \dd{t} \nonumber \\
    & V_{\ref{fig:Feynman:3p:HKK}} = -\frac{ G q_1 q_2 (m_1+m_2)}{2\pi r^2 c^2} \label{eq:3p:Potential:HKK}
\end{align}
Notice how $V_{\ref{fig:Feynman:3p:HKK}} = -2\cdot V_{\ref{fig:Feynman:Seagull:HK}}$ and will thus `destructively interfere' with one another, analogous to how $V_{\ref{fig:Feynman:3p:HHH}} = -2 \cdot V_{\ref{fig:Feynman:Seagull:HH}}$ in GR.

\section{The resulting binary Lagrangian}
The kinetic term of the binary Lagrangian is obtained from equation \eqref{eq:ppAction full} by the cross-term $\sum_a \gamma_a^{-1} m_a c^2 \times (-1)$ and expanding the Lorentz factor in powers of velocity. To 1PN this is
\begin{align}
    L_{pp}^\text{kin} & = \sum_a -m_a c^2 + \frac{1}{2} m_a v^2_a + \frac{1}{8} m_a \frac{v^4_a}{c^2} + \order{\frac{v^6}{c^4}} = -Mc^2 + \frac{1}{2}\mu v^2 + \frac{1-3\eta}{8} \mu \frac{v^4}{c^2} + \order{\frac{v^6}{c^4}}.
\end{align}
The constant term can be ignored for all dynamical purposes. In the last equality effective one-body coordinates have been implemented, where $M$ is the total mass $M=m_1+m_2$, $\mu$ is the reduced mass $\mu=\frac{m_1 m_2}{m_1+m_2}$, $\eta$ is the symmetric mass ratio $\eta=\frac{\mu}{M}$, and $v$ is the relative velocity $\vb*{v} = \dot{\vb*{r}} = \vb*{v}_1 - \vb*{v}_2$ when $\vb*{r}=\vb*{r}_1 - \vb*{r}_2$.

The contributions from standard general relativity are derived from all the diagrams \emph{only} containing graviton propagators
\begin{align}
    L_{pp}^\text{GR} = & \frac{Gm_1m_2}{r} + \frac{Gm_1m_2}{r}\left( \frac{3}{2}\frac{v_1^2+v_2^2}{c^2} - \frac{4\vb*{v}_1 \vb*{\cdot} \vb*{v}_2}{c^2} + \frac{\vb*{v}_1 \vb*{\cdot} \vb*{v}_2 - (\vb*{v}_1 \vb*{\cdot} \vb*{\hat{r}})(\vb*{v}_2 \vb*{\cdot} \vb*{\hat{r}}) }{2c^2} - \frac{G(m_1+m_2)}{2 r c^2} \right) \nonumber \\
    = & \frac{GM\mu}{r} + \frac{GM\mu}{r}\left( \frac{3}{2} \left(1-2\eta\right) \frac{ v^2 }{c^2} + \frac{4\eta v^2}{c^2} - \eta \frac{ v^2 - (\vb*{v \cdot \hat{r}})^2}{2c^2} - \frac{GM}{2rc^2} \right) +\order{\frac{v^3}{c^3}}.
\end{align}
Next the additional potentials from the inclusion of the Kalb-Ramond (effectively scalar) field
\begin{align}
	L_{pp}^\text{KR} &= \frac{q_1 q_2}{4\pi r} + \frac{q_1 q_2}{4\pi r}\left( -\frac{v_1^2+v_2^2}{2c^2} + \frac{\vb*{v}_1 \vb*{\cdot} \vb*{v}_2 - (\vb*{v}_1 \vb*{\cdot} \vb*{\hat{r}})(\vb*{v}_2 \vb*{\cdot} \vb*{\hat{r}}) }{2c^2} -\frac{\left(\frac{p_1 q_2}{q_1} + \frac{p_2 q_1}{q_2}\right)}{4\pi r} + \frac{G(m_1+m_2)}{r c^2} + \frac{G \left( \frac{m_1 q_2}{q_1} + \frac{m_2 q_1}{q_2} \right) }{r c^2}\right) \nonumber \\
    &= \frac{q_1 q_2}{4\pi r} +  \frac{q_1 q_2}{4\pi r} \left( - \frac{1-2\eta}{2}\frac{v^2}{c^2} - \eta \frac{v^2 - \left(\vb*{v \cdot \hat{r}}\right)^2}{2c^2} - \frac{p}{4 \pi r} + \frac{GM}{rc^2} + \frac{G\left( \frac{m_1 q_2}{q_1} + \frac{m_2 q_1}{q_2} \right)}{ rc^2 } \right). \label{eq:Lagragian:KR}
\end{align}
In the last line $p=\frac{p_1 q_2}{q_1} + \frac{p_2 q_1}{q_2}$. Notice that if $p_a \propto q_a$ it follows that $p\propto q_1 + q_2$, i.e. the total charge, making it analogous to how diagram \ref{fig:Feynman:Seagull:HH} is a factor of total mass correction to Newton's potential in GR ($V_\text{\ref{fig:Feynman:Seagull:HH}}=-\frac{GM\mu}{r} \cdot \frac{-GM}{2 r c^2}$).

For comparison here is the Lagrangian obtained by \citeauthor{AxionLIGO} for axions in \cite{AxionLIGO}, rewritten to have consistent notation with this paper:
\begin{align}
    L_{pp}^\phi = & \frac{ q_1 q_2 }{4\pi r} e^{-m_s r}  + \frac{ q_1 q_2 }{4\pi r} e^{-m_s r} \left( \frac{\vb*{v}_1 \vb*{\cdot} \vb*{v}_2 - (\vb*{v}_1 \vb*{\cdot} \vb*{\hat{r}})(\vb*{v}_2 \vb*{\cdot} \vb*{\hat{r}})(1+m_s r) }{2c^2} -\frac{\left(\frac{p_1 q_2}{q_1} + \frac{p_2 q_1}{q_2}\right)}{2\pi r} e^{-m_s r} - \frac{G(m_1+m_2)}{r c^2} \right) \nonumber \\
        & - \frac{G (m_1 q_2^2 + m_2 q_1^2)}{16\pi r} m_s \left[-1 + e^{-2m_s r} + 2m_s r \text{Ei}(-2m_s r) \right] + \frac{G M q_1 q_2}{2\pi rc^2} m_s \mathcal{I}(m_s r).
\end{align}